\documentclass[letterpaper]{article} 
\usepackage{aaai25}  
\usepackage{times}  
\usepackage{helvet}  
\usepackage{courier}  
\usepackage[hyphens]{url}  
\usepackage{graphicx} 
\usepackage{natbib}  
\usepackage{caption} 
\usepackage{xcolor}
\usepackage{algorithm}
\usepackage{algorithmic}

\usepackage{newfloat}
\usepackage{listings}
\DeclareCaptionStyle{ruled}{labelfont=normalfont,labelsep=colon,strut=off} 
\floatstyle{ruled}
\newfloat{listing}{tb}{lst}{}
\floatname{listing}{Listing}
\title{Anti-ESIA: Analyzing and Mitigating Impacts of Electromagnetic \\Signal Injection Attacks}

\author{
    Denglin Kang\textsuperscript{\rm 1},
    Youqian Zhang\thanks{Corresponding author: you-qian.zhang@polyu.edu.hk}\textsuperscript{\rm 2},
    Wai Cheong Tam\textsuperscript{\rm 3},
    Eugene Y. Fu\textsuperscript{\rm 2} 
}
\affiliations{
    \textsuperscript{\rm 1} University of Southern California \\
    denglink@usc.edu \\
    \textsuperscript{\rm 2} The Hong Kong Polytechnic University,
    \textsuperscript{\rm 3} National Institute of Standards and Technology
}

\begin{document}

\maketitle

\begin{abstract}
Cameras are integral components of many critical intelligent systems. 
However, a growing threat, known as Electromagnetic Signal Injection Attacks (ESIA), poses a significant risk to these systems, where ESIA enables attackers to remotely manipulate images captured by cameras, potentially leading to malicious actions and catastrophic consequences. 
Despite the severity of this threat, the underlying reasons for ESIA's effectiveness remain poorly understood, and effective countermeasures are lacking.
This paper aims to address these gaps by investigating ESIA from two distinct aspects: pixel loss and color strips. 
By analyzing these aspects separately on image classification tasks, we gain a deeper understanding of how ESIA can compromise intelligent systems. 
Additionally, we explore a lightweight solution to mitigate the effects of ESIA while acknowledging its limitations. 
Our findings provide valuable insights for future research and development in the field of camera security and intelligent systems.
\end{abstract}

\section{Introduction}
\label{sec:introduction}
As intelligent systems become increasingly integrated into our daily lives, their reliance on sensors to perceive and interact with the physical world becomes paramount. 
Cameras, in particular, play a critical role in providing these systems with visual capabilities, enabling them to ``see'' and ``understand'' their surroundings. 
From autonomous vehicles to robotic systems, countless applications rely on the integrity of camera-captured images to make safety- or security-critical decisions.

Unfortunately, an emerging threat has emerged: Electromagnetic Signal Injection Attacks (ESIA)~\cite{jiang23glitchhiker, zhang2024esia}. 
Unlike traditional cyberattacks that target software vulnerabilities, ESIA operates at the physical layer, exploiting hardware imperfections to inject adversarial electromagnetic signals into camera circuits wirelessly, and as such disrupt the transmission of pixel information.
Thus, attackers can remotely manipulate captured images, potentially leading to catastrophic consequences.
For instance, previous studies~\cite{jiang23glitchhiker, zhang2024modeling, zhang2024esia} showed that ESIA could tamper with cameras of autonomous vehicles, and cause their perception system to fail to detect obstacles in front, leading to fatal collisions.

While ESIA has not yet been widely abused by attackers, its potential for causing significant harm is undeniable. 
Understanding the mechanisms through which ESIA can impact AI based systems is crucial for developing effective countermeasures. 
However, research in this area remains limited, leaving billions of systems vulnerable to these attacks.
This paper aims to address these knowledge gaps by making the following contributions:
\begin{itemize}
    \item This is the first work that identifies and demonstrates two primary factors that contribute to the degradation of intelligent systems. 
    \item This work is the first that introduces a practical and efficient method to mitigate the effects of ESIA.
\end{itemize}

\section{Analysis of Synthetic Attack Impacts}
\label{sec:modeling}
\begin{figure}[t]
\centering
\includegraphics[width=0.475\textwidth]{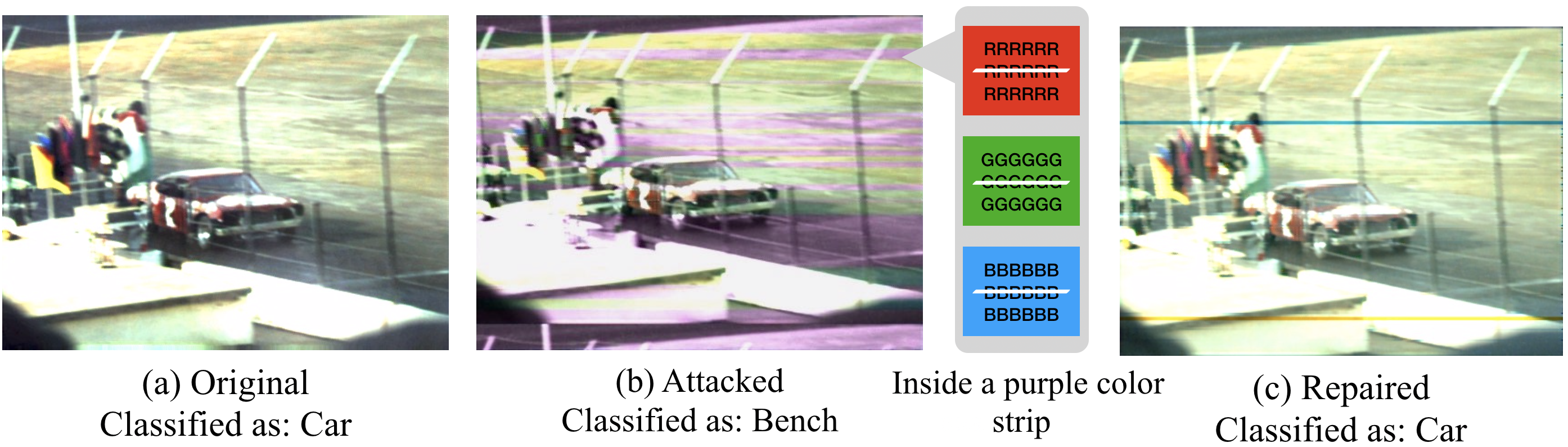}
\caption{ESIA can cause pixel loss and color strips, misleading intelligent systems to make wrong decisions.}
\label{fig:color_strip}
\end{figure}

The impact of ESIA on computer vision systems often manifests visually as color strips across the image, as shown in Figure~\ref{fig:color_strip}. 
This phenomenon occurs due to the attack causing certain rows of pixels to be disregarded by the system during image processing. 
When the system attempts to reconstruct the image, the missing rows are replaced by subsequent rows, causing incorrect color interpretation. 
As a result, this replacement results in visible color strips. 
For a detailed explanation of the attack mechanism, readers are referred to previous work~\cite{jiang23glitchhiker, zhang2024modeling}.
Upon analyzing the effects of such attacks, we identify two key contributing factors: pixel loss, and color strips. 
These factors together cause significant degradation in the performance of artificial intelligence (AI) models that rely on images.

Let's define a strictly increasing tuple $R = (r_0, r_1, ..., r_{n-1}),~n \in Z^{+}$, representing the dropped rows of pixels. 
A color strip is caused by a pair of dropped rows, i.e., $r_i$ and $r_{i+1}$, where $i$ is even.
Note that if $(n - 1)$ is an even number, it means the last strip will end at the bottom of the image.
The overall degradation in model performance due to the attack can be expressed as:
\[
    Degradation = f(R) + g(R)
\]
where $f(\cdot)$ is a function of the pixel loss, and $g(\cdot)$ is the function of the color strips.

We randomly picked 495 photos from the validation set from ImageNet Large Scale Visual Recognition Challenge datasets~\cite{ILSVRC15}.
We use the synthetic method proposed by~\cite{zhang2024modeling} to generate attack images without color strips, where $n$ is set to 0, 16, 22, 46, 66, and 76, as they correspond to 0\%, 10\%, 15\%, 30\%, 45\% and 50\% of the total number of rows of the images.
The positions of the dropped rows are randomly chosen.
Then, we generate attack images with color strips similarly.
Despite that there are many different computer vision tasks, we pick image classification, as it is one of the most fundamental ones; we use the model vit-b-32~\cite{dosovitskiy2020image}, which is one of the commonly used and cutting-edge methods.

\begin{figure}[t]
\centering
\includegraphics[width=0.4\textwidth]{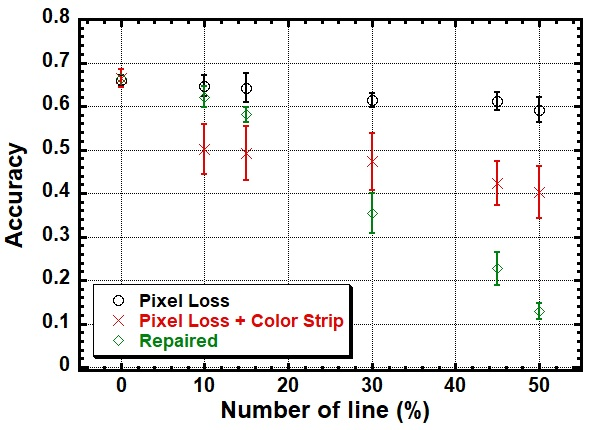}
\caption{Performance is degraded by ESIA with or without color strips, and performance can be recovered after applying our mitigation method.}
\label{fig:attack_impacts_w_wo_color_strips}
\end{figure}

We use accuracy as the metric for image classification, and the results are presented in Figure~\ref{fig:attack_impacts_w_wo_color_strips}. 
It demonstrates that while pixel loss alone leads to a moderate performance decline, the combination of pixel loss and color strips causes a significantly more severe degradation. 
Specifically, color strips contribute to approximately 74\% of the overall performance drop.

\section{Mitigation Method}
\label{sec:mitigation}
To address the impact of ESIA, we explore a simple yet effective mitigation strategy based on median interpolation. For each dropped pixel, we examine its neighboring 3x3 grid and select the median value as a replacement. This approach leverages the spatial correlation of pixel values to estimate missing data.

Our experiments demonstrate that this method can partially recover performance for moderate levels of pixel loss and color distortion, as shown in Figure~\ref{fig:attack_impacts_w_wo_color_strips}. 
However, as the number of dropped rows increases, the effectiveness of median interpolation diminishes. This limitation arises from the inherent difficulty of accurately reconstructing missing data when substantial information is lost.

\section{Conclusion and Future Work}
This paper has investigated the impact of Electromagnetic Signal Injection Attacks (ESIA) on intelligent systems that rely on camera-captured images. By analyzing the effects of ESIA, we have identified two primary factors contributing to model degradation: pixel loss and color strips. 
Our findings demonstrate that color distortion has a more significant impact on model performance than pixel loss.
To mitigate the effects of ESIA, we explore a lightweight solution based on median interpolation. While this method can effectively address some instances of color distortion, it exhibits limitations when dealing with severe attacks that result in the loss of a large number of pixels.
Future research could explore more advanced mitigation techniques to address the limitations of the proposed method. 
For example, leveraging deep learning models to predict and correct missing pixel values or exploring techniques to detect and remove color strips more accurately could be promising avenues.
Additionally, investigating the impact of ESIA on other types of intelligent systems, such as those used in healthcare or critical infrastructure, is essential. 
Understanding the vulnerabilities of these systems will make sure they provide reliable services to people.

\bibliography{aaai25}

\begin{thebibliography}{5}
\providecommand{\natexlab}[1]{#1}

\bibitem[{Dosovitskiy(2020)}]{dosovitskiy2020image}
Dosovitskiy, A. 2020.
\newblock {An Image Is Worth 16x16 Words: Transformers for Image Recognition at Scale}.
\newblock \emph{arXiv preprint arXiv:2010.11929}.

\bibitem[{Jiang et~al.(2023)Jiang, Ji, Yan, Xie, Lou, and Xu}]{jiang23glitchhiker}
Jiang, Q.; Ji, X.; Yan, C.; Xie, Z.; Lou, H.; and Xu, W. 2023.
\newblock {GlitchHiker: Uncovering Vulnerabilities of Image Signal Transmission with IEMI}.
\newblock In \emph{The 32nd USENIX Security Symposium}.

\bibitem[{Russakovsky et~al.(2015)Russakovsky, Deng, Su, Krause, Satheesh, Ma, Huang, Karpathy, Khosla, Bernstein, Berg, and Fei-Fei}]{ILSVRC15}
Russakovsky, O.; Deng, J.; Su, H.; Krause, J.; Satheesh, S.; Ma, S.; Huang, Z.; Karpathy, A.; Khosla, A.; Bernstein, M.; Berg, A.~C.; and Fei-Fei, L. 2015.
\newblock {ImageNet Large Scale Visual Recognition Challenge}.
\newblock \emph{International Journal of Computer Vision (IJCV)}, 115(3): 211--252.

\bibitem[{Zhang et~al.(2024{\natexlab{a}})Zhang, Cheung, Yang, Zhai, Shen, Ji, Fu, Chau, and Luo}]{zhang2024modeling}
Zhang, Y.; Cheung, M.; Yang, C.; Zhai, X.; Shen, Z.; Ji, X.; Fu, E.~Y.; Chau, S.-Y.; and Luo, X. 2024{\natexlab{a}}.
\newblock {Modeling Electromagnetic Signal Injection Attacks on Camera-based Smart Systems: Applications and Mitigation}.
\newblock \emph{arXiv preprint arXiv:2408.05124}.

\bibitem[{Zhang et~al.(2024{\natexlab{b}})Zhang, Yang, Fu, Jiang, Yan, Chau, Ngai, Leong, Luo, and Xu}]{zhang2024esia}
Zhang, Y.; Yang, C.; Fu, Y.; Jiang, Q.; Yan, C.; Chau, S.-Y.; Ngai, G.; Leong, H.-v.; Luo, X.; and Xu, W. 2024{\natexlab{b}}.
\newblock {Understanding Impacts of Electromagnetic Signal Injection Attacks on Object Detection}.
\newblock In \emph{IEEE International Conference on Multimedia and Expo}. IEEE.

\end{thebibliography}

\end{document}